\documentclass[twocolumn,showpacs,aps,prl,floats]{revtex4}
\usepackage{graphicx,bm,dcolumn}
\usepackage{amssymb}

\input epsf.tex
\include{bibliohead}

\def\plotfull#1{\centering \leavevmode
\epsfxsize= 1.75\columnwidth \epsfbox{#1}}

\def\plotthree#1#2#3{\centering \leavevmode
\epsfxsize=.6\columnwidth \epsfbox{#1}
\epsfxsize=.75\columnwidth \epsfbox{#2} 
\epsfxsize=.6\columnwidth \epsfbox{#3}}

\def\lsim{\mathrel{\rlap{\lower4pt\hbox{\hskip1pt$\sim$}}
    \raise1pt\hbox{$<$}}}                
\def\gsim{\mathrel{\rlap{\lower4pt\hbox{\hskip1pt$\sim$}}
    \raise1pt\hbox{$>$}}}                

\begin{document}
\preprint{IUCAA}

\title{Statistical Isotropy of CMB and Cosmic Topology}
\author{Amir Hajian and Tarun Souradeep } 
\affiliation{Inter-University Centre for Astronomy and Astrophysics,\\
  Post Bag 4, Ganeshkhind, Pune 411007, India}

\begin{abstract}

  
  The breakdown of statistical homogeneity and isotropy of cosmic
  perturbations is a generic feature of ultra large scale structure of
  the cosmos, in particular, of non trivial cosmic topology. The
  statistical isotropy (SI) of the Cosmic Microwave Background
  temperature fluctuations (CMB anisotropy) is sensitive to this
  breakdown on the largest scales comparable to, and even beyond the
  cosmic horizon.  We study a set of measures, $\kappa_\ell$
  ($\ell=1,2,3, \ldots$) which for non-zero values indicate and
  quantify statistical isotropy violations in a CMB map.  The main
  goal here is to interpret the $\kappa_\ell$ spectrum and relate it
  to characteristic patterns in the correlation function of CMB
  anisotropy arising from cosmic topology.  We numerically compute the
  predicted $\kappa_\ell$ spectrum for CMB anisotropy in flat torus
  universe models. The essential features are captured in the leading
  order approximation to the correlation function where $\kappa_\ell$
  can be calculated analytically.  The $\kappa_\ell$ spectrum is shown
  to reflect the number, importance and relative orientation of
  principal directions in the CMB correlation dictated by the shape of
  the Dirichlet domain (DD) of the compact space and its size relative
  to cosmic horizon. Hence, besides detecting cosmic topology,
  $\kappa_\ell$ can discriminate between different topology of the
  universe complementing ongoing search for cosmic topology in CMB
  anisotropy data.
 
\end{abstract}

\maketitle

In standard cosmology, the Cosmic Microwave Background (CMB)
anisotropy is expected to be statistically isotropic, i.e.,
statistical expectation values of the temperature fluctuations $\Delta
T(\hat q)$ are preserved under rotations of the sky. In particular,
the angular correlation function $C(\hat{q},\,
\hat{q}^\prime)\equiv\langle\Delta T(\hat q)\Delta T(\hat
q^\prime)\rangle$ is rotationally invariant for Gaussian fields. In
spherical harmonic space, where $\Delta T(\hat q)= \sum_{lm}a_{lm}
Y_{lm}(\hat q)$ this translates to a diagonal $\langle a_{lm}
a^*_{l^\prime m^\prime}\rangle=C_{l}
\delta_{ll^\prime}\delta_{mm^\prime}$ where $C_l$ is the widely used
angular power spectrum of CMB anisotropy.

It is important to determine whether the CMB sky is a realization of a
statistically isotropic process, or not from the observations
themselves. We study a set of measures $\kappa_\ell$ ($\ell=1,2,3,
\ldots$) that measure violation of statistical isotropy~\cite{us_apj}.

The detection of statistical isotropy (SI) violations can have
exciting and far-reaching implication for cosmology.  The realization
that the universe with the same local geometry has many different
choices of global topology has been a theoretical curiosity as old as
modern cosmology. Motivations for cosmic topology and their
consequences have been extensively studied~\cite{costop}. CMB
anisotropy measurements have brought cosmic topology from the realm of
theoretical possibility to within the grasp of
observations~\cite{costop,bps}.  {\em A generic consequence of cosmic
  topology is the breaking of statistical isotropy in characteristic
  patterns determined by the photon geodesic structure of the
  manifold.}  Global isotropy of space is violated in all multi
connected models (except $S^3/Z_2$).  In cosmology, the Dirichlet
domain (DD) constructed around the observer represents the universe as
`seen' by the observer.  The SI breakdown is apparent in the principal
axes present in the shape of the DD constructed with the observer
located at the basepoint.

In this paper we compute and study the $\kappa_\ell$ spectrum of SI
violation arising in flat (Euclidean) simple torus models with a
cubic, cuboidal and more generally, parallelepiped (squeezed)
fundamental domain. The CMB anisotropy in torus spaces has been well
studied~\cite{tor_refs,bow_fer02}.  We can relate the $\kappa_\ell$
spectrum to the principal directions normal to pair of faces of the
DD, their relative orientation and the relative importance given by
the distance to the faces along them.  Along the most dominant axes,
the distance is minimum, and equals the {\em inradius}, $R_<$, the
radii of largest sphere fully enclosed within the DD~\cite{bps}.

\begin{figure*}[t]
\plotthree{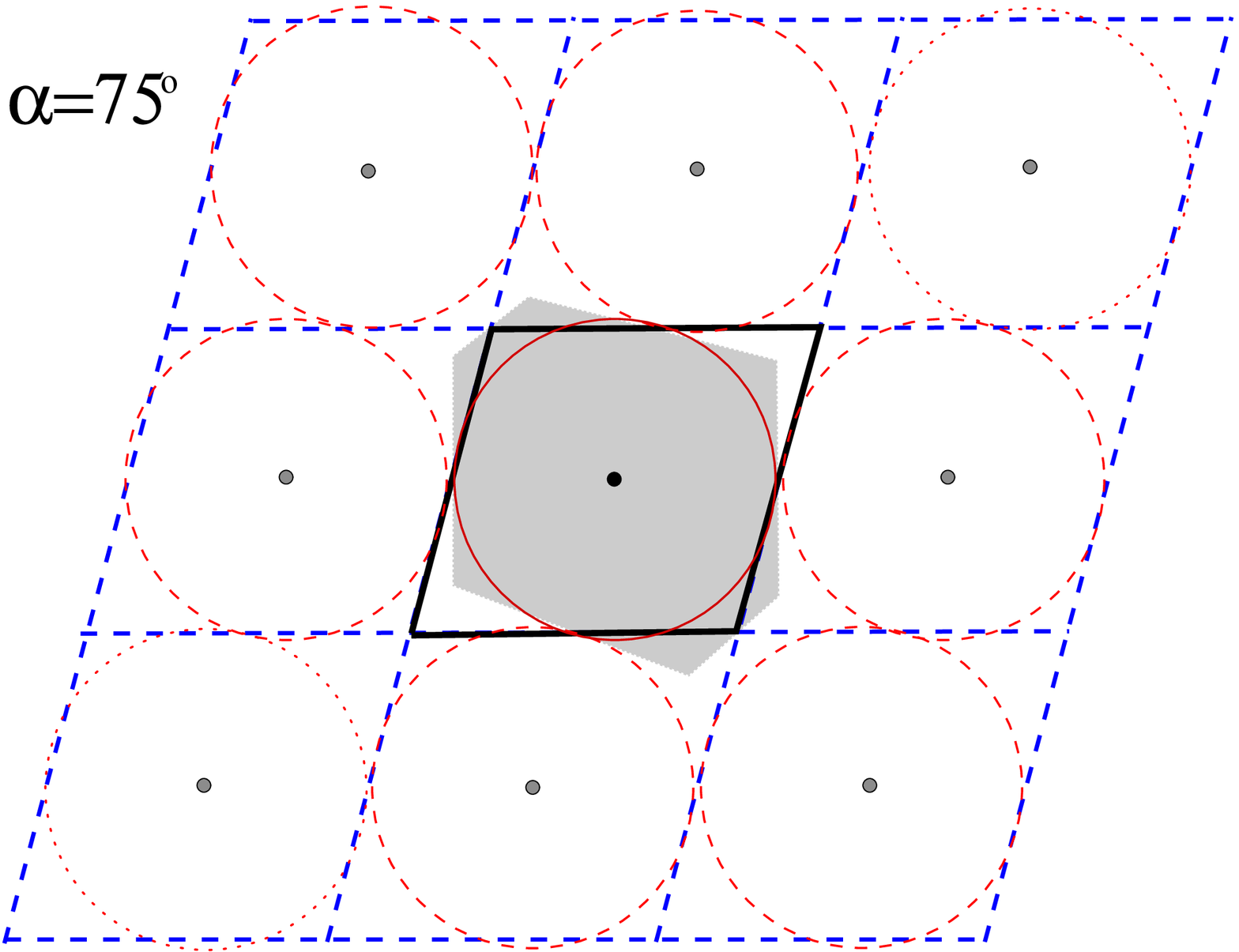}{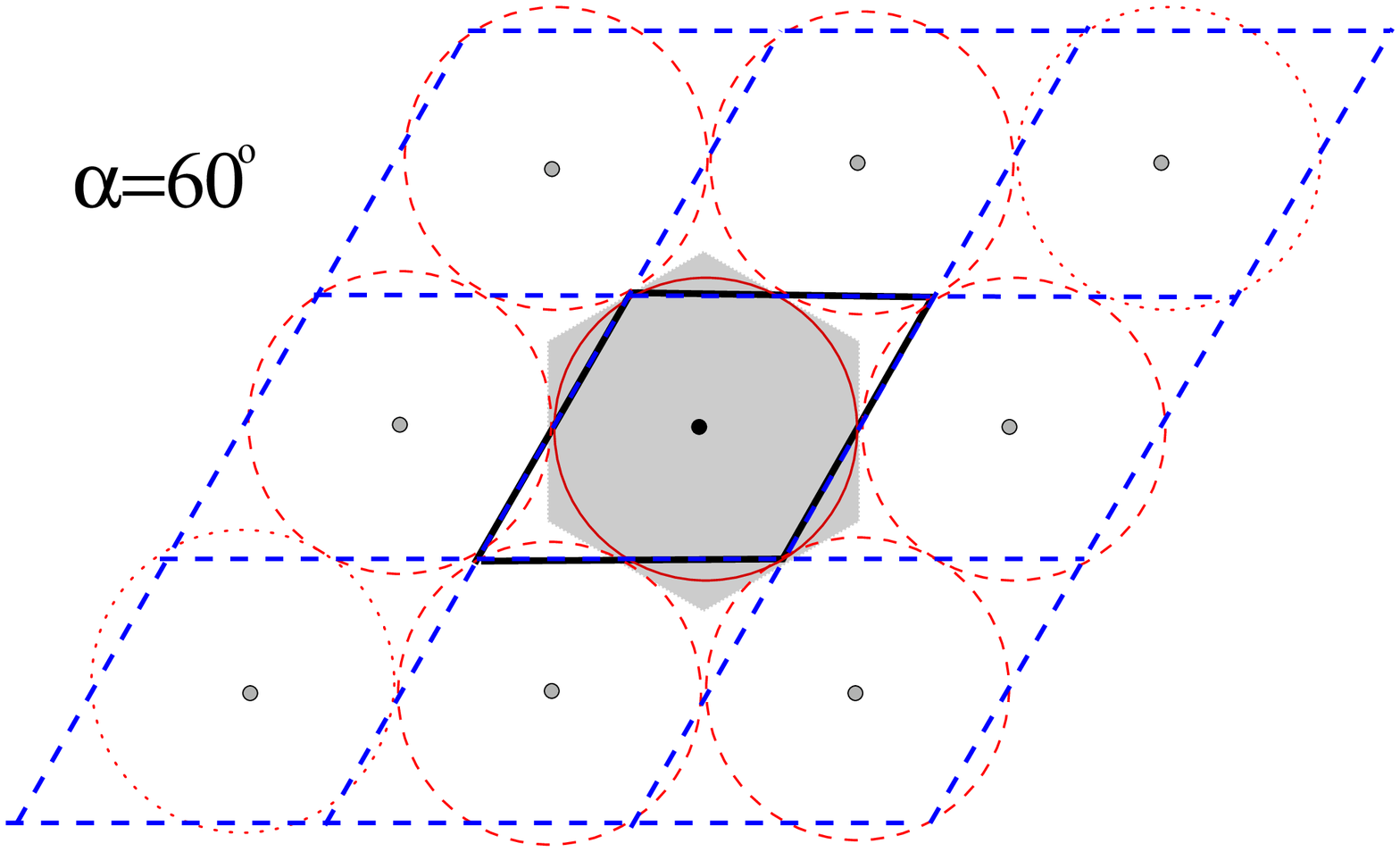}{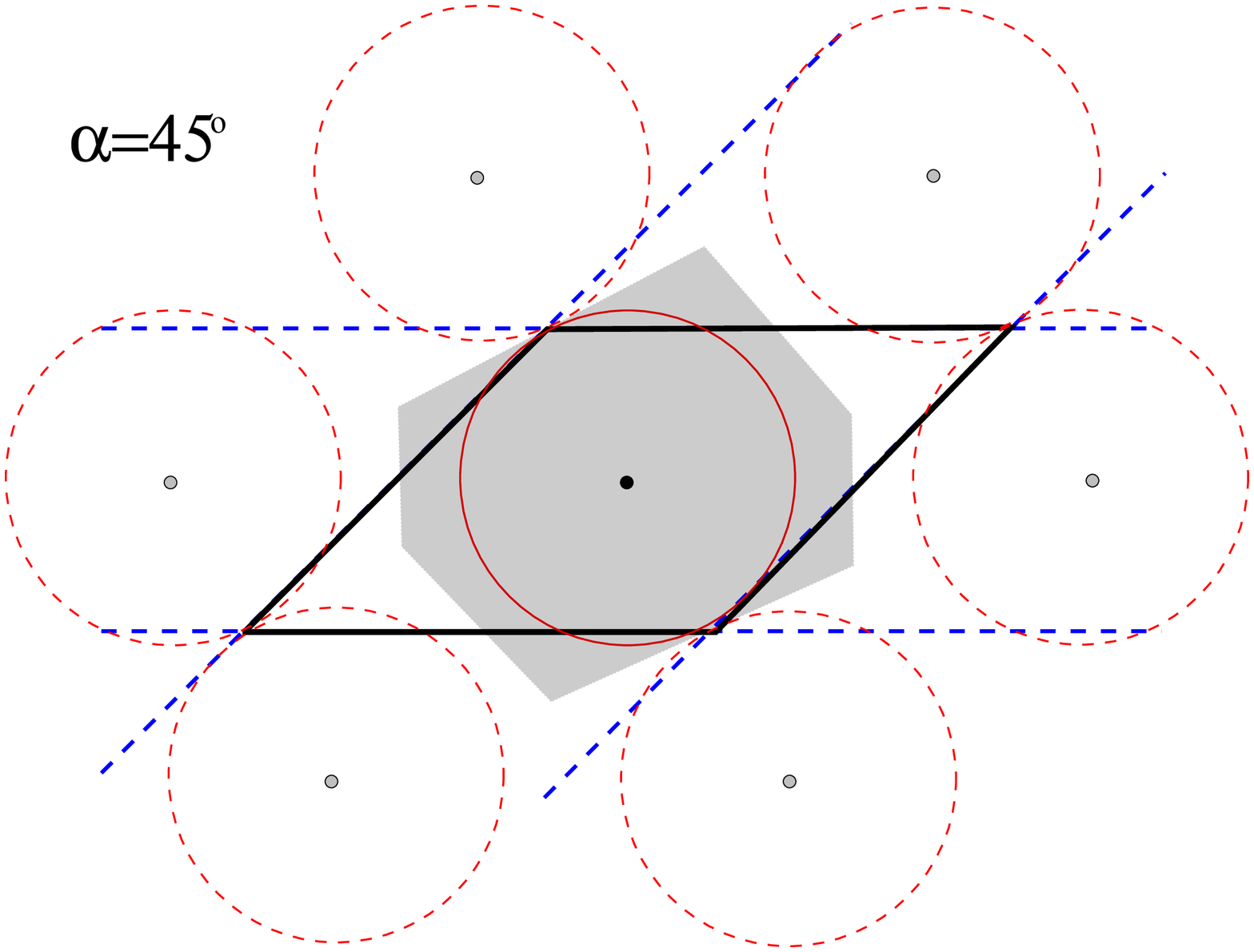}
  \caption{The pattern in CMB correlation in a multi-connected  universe, 
    ${\cal M}\equiv{\cal M}^u/\Gamma$ is related to the distribution
    of `images' of the sphere of last scattering (SLS) on the
    universal cover. The three figures illustrate this for three
    values of $\alpha=45^\circ, 60^\circ,75^\circ$ in a squeezed
    torus. The solid parallelepiped is the fundamental domain and the
    dashed show its images tessellate,${\cal M}^u$. The solid circle
    is the SLS and dashed/dotted ones are its images. In each case,
    the radius of SLS is equal to the inradius, $R_<$ of ${\cal M}$.
    The shaded polygon is the Dirichlet domain (DD). The closest SLS
    images which determine the DD are dashed (others are dotted). Note
    the hexagonal shape of DD when $\alpha\neq\pi/2$ and is equal
    sided for $\alpha=\pi/3$. For smaller $\alpha$, the DD
    approximates an elongated cuboid.}
  \label{sqtor}
\end{figure*}

The set of measures $\kappa_\ell$ of statistical isotropy violation is
defined as
\begin{equation}\label{kl}
 \kappa_\ell= \int\!\!  d\Omega\!\!\int\!\!  d\Omega^\prime 
 \left[\frac{(2\ell+1)}{8\pi^2}\!\!\int d{\mathcal R} \chi_\ell({\mathcal R})
C({\mathcal R}\hat{q},{\mathcal R}\hat{q}^\prime)\right]^2\!\!\!\!,
\end{equation}
where $ C({\mathcal R}\hat{q},\, {\mathcal R}\hat{q}^\prime)$ is the
two point correlation between ${\mathcal R}\hat{q}\,$ and $ {\mathcal
  R}\hat{q}^\prime$ obtained by rotating $\hat{q}$ and
$\hat{q}^\prime$ by an element ${\mathcal R}$ of the rotation
group~\cite{us_apj}.  The measures $\kappa_\ell$ involve angular
average of the correlation weighed by the characteristic function of
the rotation group $ \chi_\ell({\mathcal R})=\sum_{M}
D_{MM}^{\ell}({\mathcal R})$ where $ D_{MM^\prime}^{\ell}$ are the
Wigner D-functions~\cite{Var}.  When SI holds $ C({\mathcal
  R}\hat{q},\, {\mathcal R}\hat{q}^\prime)\,=\,C(\hat{q},\,
\hat{q}^\prime)$ is invariant under rotation, and eq.~(\ref{kl}) gives
$\kappa_\ell = \kappa_0\, \delta_{\ell 0}$ due to the orthonormality
of $\chi_\ell({\mathcal R})$.  {\em Hence, non-zero $\kappa_\ell$ for
  $\ell >0$ measure violation of statistical isotropy.}
 
The measure $\kappa_\ell$ has a clear interpretation in harmonic
space.  The two point correlation $C(\hat{q},\, \hat{q}^\prime)$ can
be expanded in terms of the orthonormal set of bipolar spherical
harmonics as
\begin{equation}\label{bipolar}
 C(\hat{q},\, \hat{q}^\prime)\, =\, \sum_{ll^\prime\ell M} A_{ll^\prime}^{\ell M}
 \{Y_{l}(\hat{q}) \otimes Y_{l^\prime}(\hat{q}^\prime)\}_{\ell M}\,,
\end{equation}
where $A_{ll^\prime}^{\ell M}$ are the coefficients of the expansion.
These coefficients are related to `angular momentum' sum over the
covariances $\langle a_{lm}a^*_{l^\prime m^\prime}\rangle$ as 
\begin{equation}\label{alml1l2}
 A_{ll^\prime}^{\ell M} = 
\sum_{m m^\prime} \langle a_{lm}a^*_{l^\prime m^\prime}\rangle
\, \, (-1)^{m^\prime} \mathfrak{ C}^{\ell M}_{lml^\prime -m^\prime}\,,
\end{equation}
where $\mathfrak{C}^{\ell M}_{lml^\prime m^\prime}$ are Clebsch-Gordan
coefficients~\cite{Var}.  When SI holds $\langle a_{lm}a^*_{l^\prime
  m^\prime}\rangle=C_{l}\delta_{ll^\prime}\delta_{mm^\prime}$,
implying $A_{ll^\prime}^{\ell M}=(-1)^l C_{l} (2l+1)^{1/2} \,
\delta_{ll^\prime}\, \delta_{\ell 0}\, \delta_{M0}$. $A_{ll}^{00}$
represent the statistically isotropic part of a general correlation
function. The bipolar functions transform just like ordinary spherical
harmonic function $Y_{LM}$ under rotation~\cite{Var}.  Substituting
the expansion eq.~(\ref{bipolar}) into eq.~(\ref{kl}) we can show that
$\kappa_\ell =\sum_{ll^\prime M} |A_{ll^\prime}^{\ell M}|^2$ is positive
semidefinite and is also given by
\begin{equation}
 \kappa_\ell = \frac{2\ell+1}{8\pi^2}\!\!
\int\!\! d{\mathcal R}\chi_\ell({\mathcal R})
\sum_{lml^\prime m^\prime}\langle a_{lm}a^*_{l^\prime m^\prime}
\rangle\langle a_{lm}a^*_{l^\prime m^\prime}
\rangle^{\mathcal R}\!\!,
\end{equation}
where $\langle\ldots\rangle^{\mathcal R}$ is computed in a frame
rotated by ${\mathcal R}$~\cite{us_apj}.

The compact spaces with Euclidean geometry (zero curvature) have been
completely classified. In three dimensions, there are known to be six
possible topologies that lead to orientable spaces
~\cite{costop,wol94vin93}.  The simple flat torus, ${\cal M} = T^3$,
is obtained by identifying the universal cover ${\cal M}^u={\cal E}^3$
under a discrete group of translations along three non-degenerate
axes, ${\mathbf s_1,\mathbf s_2,\mathbf s_3}$: $ {\mathbf s_i} \to
{\mathbf s_i} + {\mathbf n} L_i$, where $L_i$ is the identification
length of the torus along $s_i$ and ${\bf n}$ is a vector with integer
components.  In the most general form, the fundamental domain (FD) is
a parallelepiped defined by three sides $L_i$ and the three angles
$\alpha_i$ between the axes (We call it squeezed torus).  If ${\mathbf
s_i}$ are orthogonal then one gets cuboidal FD, which for equal $L_i$
reduces to the cubic torus.  The cuboid and squeezed spaces which can
be obtained by a linear coordinate transformation ${\cal L}$ on cubic
torus~\cite{bow_fer02} can have distinctly different global
symmetry~\footnote{For cubic torus the Dirichlet domain (DD) matches
the fundamental domain (FD).  However, for torus spaces with cuboid
and parallelepiped FD, the corresponding DD is very different, e.g.,
hexagonal prism (see Fig~\ref{sqtor}).}.

\begin{figure*}
\plotfull{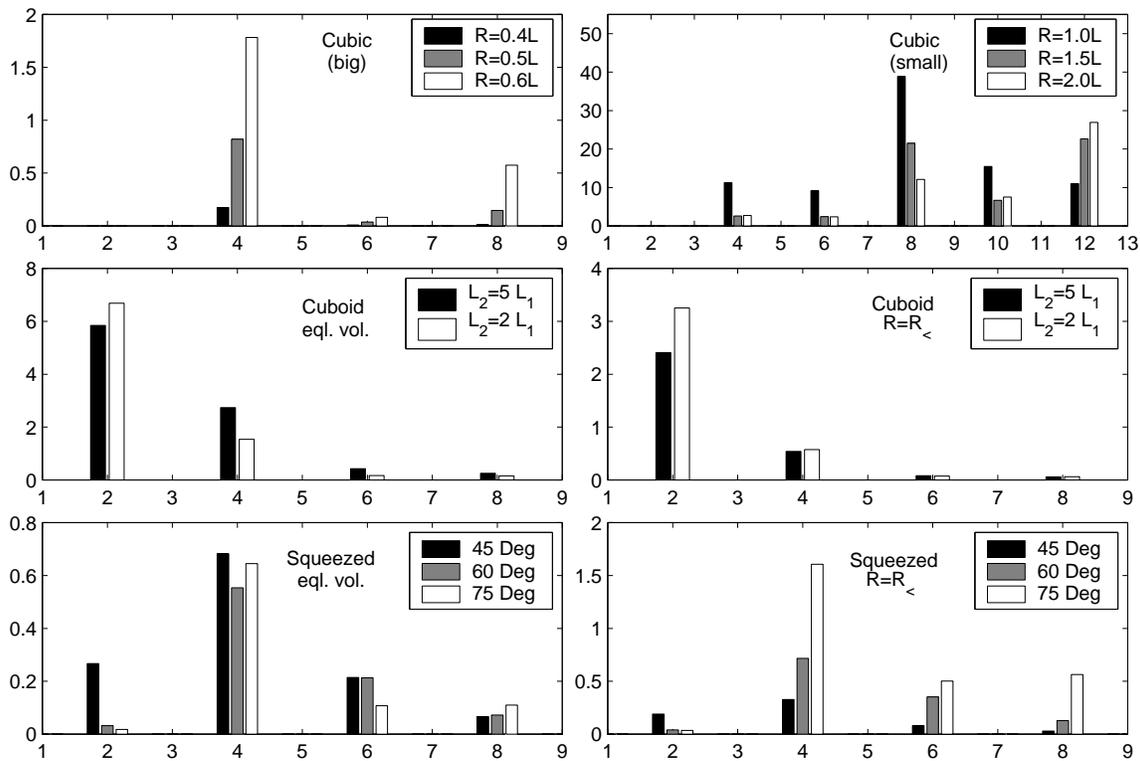}
  \caption{The $\kappa_{\ell}$ spectra for flat tori models are plotted.
    The top row panels are for cubic tori spaces. The left panel shows
    spaces of volume, $V_{\cal M}$, larger than the volume $V_{*}$
    contained in the sphere of last scattering (SLS) with $V_{\cal
      M}/V_{*} = 3.7,1.9,1.1$, respectively.  The right panel shows
    small spaces with $V_{\cal M}/V_{*} = 0.24,0.07,0.03$,
    respectively.  Note that $\kappa_2=0$ for cubic tori.  The middle
    panels consider cuboid tori with $1:5$ and $1:2$ ratio of
    identification lengths. The bottom panels show $\kappa_{\ell}$ for
    equal-sided squeezed tori with $\alpha=45^\circ,60^\circ$ and
    $75^\circ$. In the middle and bottom rows, the right panels show
    the case when radius of SLS, $R_*=R_<$ the inradius of the space.
    Here, the SLS just touches its nearest images (see
    Fig.~\ref{sqtor}) which is at the threshold where CMB anisotropy
    is multiply imaged for larger $R_*$. The cases in the left panels
    of lower two rows have $V_{\cal M}/V_{*} =1$ and are at the divide
    between large and small spaces. }
  \label{kappaltorus}
\end{figure*}

We restrict attention to the case where CMB anisotropy arises entirely
at the sphere of last scattering (SLS) of radius $R_{*}$
(nearly equal to observable horizon).  Invoking method of images, the
CMB correlation pattern on the SLS is known to be dictated by the
distribution of nearest `images' of the SLS on the universal
cover~\cite{bps}. The correlations are distorted even when the SLS and
its images do not intersect ($R_* < R_<$). When SLS intersects its
images the CMB sky is multiply imaged in characteristic correlation
pattern of pairs of circles~\cite{circles}.  Fig.~\ref{sqtor}
illustrates the role of the nearest SLS images and related DD in
defining the principal directions in the correlation for a few cases.

We compute $C({\hat q,\hat q^\prime})$ for CMB anisotropy in torus
space using regularized method of images~\cite{bps}. We can compute
the $\kappa_\ell$ in real space using eq.~(\ref{kl}) or from
$A_{ll^\prime}^{\ell M}$ using eq.~(\ref{alml1l2}) in harmonic space.
Fig.~\ref{kappaltorus} plots the $\kappa_\ell$ spectrum for a number
of cubic, cuboidal and squeezed torus spaces. We find the following
interesting results~:

{\bf i.} $\kappa_\ell=0$ for odd $\ell$ for all torus models. This
does not hold for compact space of non-zero curvature, e.g., compact
hyperbolic spaces.

{\bf ii.}  For cubic torus $\kappa_2=0$. $\kappa_2$ is non-zero for cuboidal 
and squeezed torus. This is a clear signature of non-cubic torus where the 
DD differs from the FD and has more than three principle axes.

{\bf iii.} For equal-sided squeezed torus, $\kappa_4$ , decreases as
$\alpha$ decreases from $90^\circ$ to $60^\circ$ as $R_<$ increases.
For $\alpha<60^\circ$ sharply increases with decreasing $\alpha$ as
$R_<$ decreases sharply~\cite{bow_fer02}.

{\bf iv.} $\kappa_2=0$ increases monotonically as $\alpha$ decreases
from $90^\circ$. The trend is well fit by eq.~(\ref{kasqt3}).

{\bf v.} The peak of $\kappa_\ell$ shifts to larger $\ell$ for small
spaces.

\noindent The results can be understood using  the
leading order terms of the correlation function in a torus where
$\kappa_\ell$ can be calculated analytically. For brevity, we outline
the steps for the cubic torus. The results for cuboid and squeezed
case is readily obtained using the transformation ${\mathcal L}$.
Here we list only the results leaving details of the calculation to a
more comprehensive publication~\cite{us_inprep}.

The spatial correlation function of gravitational potential in the
periodic box implied by the $T^3$ topology with cubic fundamental
domain is
\begin{equation}
\xi_\Phi({\bf x,x^\prime}) = L^{-3} \sum _{{\bf n}}
P_\Phi(k_{\bf n}) \,\,{\mathrm e}^{-i \frac{2 \pi {\bf n}}{L}\cdot ({\bf x} - 
{\bf x}^\prime)},
\label{xic_tor}
\end{equation}
where ${\bf n}$ is 3-tuple of integers, $k^2_{\bf n} = (2\pi/L)^2
({\bf n} \cdot {\bf n})$, and the term with ${\bf n}\cdot{\bf n}=0$ is
excluded from the summation.  For Naive Sachs-Wolfe CMB anisotropy
arising at the SLS ( $\Delta T = 1/3\Phi $), the correlation function,
$C(\hat q, \hat q^\prime)$ is given by the spatial correlation of
$\Phi$ at points on the SLS along the two directions $\hat q$ and
$\hat q^\prime$ as
\begin{equation}
C({\hat q,\hat q^\prime}) = L^{-3} \sum _{{\bf n}}
P_\Phi(k_{\bf n}) \,\,{\mathrm e}^{-i \pi 
(\epsilon_{\hat q} {\bf n}\cdot {\hat q} - \epsilon_{\hat q^\prime} {\bf n}\cdot 
{\hat q}^\prime)},
\label{C_tor}
\end{equation}
where the small parameter $\epsilon_{\hat q} \le 1 $ is the physical
distance to the SLS along $\hat q$ in units of $L/2$ (more generally,
$\bar L/2$ where $\bar L= (L_1L_2L_3)^{1/3}$).  Contribution to
$C({\hat q,\hat q^\prime})$ from large wavenumbers, $|\bf n|\epsilon
\gg 1$ is expected to be approximately statistically isotropic.  The
SI violation is found in the low wavenumbers.

When the SLS is contained with the fundamental domain around the
observer, i.e., the CMB anisotropy is not multiply imaged on the sky,
$\epsilon = 2R_{*}/L$ is a constant.  When $\epsilon$ is a small
constant, the leading order terms in the correlation function
eq.~(\ref{C_tor}) can be readily obtained in power series expansion in
powers of $\epsilon$.  For the lowest wavenumbers $|{\mathbf n}|^2=1$
in a cuboid FD torus
\begin{equation}
C({\hat q,\hat q^\prime}) \approx 2 \sum_i P_\Phi({2\pi}/{L_i}) 
\cos(\pi\epsilon\beta_i\Delta q_i).
\label{appcorr}
\end{equation}
where $\Delta q_i$ are the components of ${\mathbf \Delta q} =\hat
q-\hat q^\prime$ along the three axes of the torus and $\beta_i = \bar
L /L_i$.  Only even powers of $\epsilon \Delta q_i$ are present in the
expansion eq.~(\ref {appcorr}). This holds for the terms from higher
wavenumbers ($|{\mathbf n}|^2=2$ or $3$) and explains the strictly
zero $\kappa_\ell$ for odd $\ell$~\cite{us_inprep}.

In a cubic (equal sided) torus, up to the leading order SI violating
term, the correlation is
\begin{equation}
C({\hat q,\hat q^\prime})  \approx C_0 \left[1 - \epsilon^2\, 
|{\mathbf \Delta q}|^2 + 3\,\epsilon^4 \, \sum_{i=1}^3
(\Delta q_i)^4   \right].
\label{acubict3}
\end{equation}
We retain the $\epsilon^4$ term since the term at $\epsilon^2$ is
explicitly rotationally invariant, hence does not contribute to the
violation of SI. The non-zero $\kappa_\ell$ can be analytically computed
to be
\begin{eqnarray}
\frac{\kappa_0}{C_0^2}\, &=&\,\pi^2(1-4\epsilon^2 
+\frac{368}{15} \epsilon^4-\frac{288}{5}\epsilon^6+\frac{20736}{125}\epsilon^8)
\nonumber \\
\frac{\kappa_4}{C_0^2}\, &=&\, \frac{12288 \pi^2 }{875} \epsilon^8  
\end{eqnarray}
The first non zero $\kappa_\ell$ in cubic (equal-sided) torus occurs
at $\ell=4$. $\kappa_4 \sim \epsilon^8$ falls off rapidly as $\epsilon
\rightarrow 0$ for large spaces.

For the cuboidal (unequal-sided) torus, the correlation violates SI at
order $\epsilon^2$
\begin{equation}
C({\hat q,\hat q^\prime}) \approx  C_0 \left[1 - \epsilon^2 \sum_{i=1}^3
\beta_i^2\,(\Delta q_i)^2 \right].
\label{acubiodt3}
\end{equation}
The non zero $\kappa_\ell$ corresponding to the correlation
eq.~(\ref{acubiodt3}) are
\begin{eqnarray}
\frac{\kappa_0}{C_0^2}\, &=&\, \pi^2(1-\frac{4}{3} \beta^2 \epsilon^2+\frac{16}{27} \beta^4\epsilon^4)
\nonumber \\
\frac{\kappa_2}{C_0^2}\, &=&\, \frac{64\pi^2}{135}\left[\beta^4-3 (\beta_1^2\beta_2^2+\beta_1^2\beta_3^2+\beta_2^2\beta_3^2)\right]\epsilon^4 
\label{kacuboidt33}
\end{eqnarray}
where $\beta^2=\sum_{i=1}^3 \beta_i^2$. The $\kappa_2\sim\epsilon^4$
signal falls off slower than the $\kappa_4$ signal in cubic space for
large spaces ($\epsilon \rightarrow 0$).

For the squeezed torus, the linear transformation can be applied to
obtain the expansion up to the leading order SI violating term.  For
simplicity, we restrict to equal sided squeezed torus with one
non-orthogonal pair of axes (${\mathbf s_1}\cdot{\mathbf
  s_2}=\cos\alpha$). The leading order approximation to correlation
$C({\hat q,\hat q^\prime})$ has the form

\begin{equation}
 C_0 \left[1 - \epsilon^2 
 \frac{|{\mathbf \Delta q}|^2 + 2\cos{\alpha} \Delta q_1\Delta q_2+\cos^2{\alpha}\Delta q_3^2)}{3 \sin^2{\alpha}}  \right].
\label{asqt3}
\end{equation}
The corresponding  $\kappa_{\ell}$ spectrum is 
\begin{eqnarray}
\frac{\kappa_0}{C_0^2} &=&\pi^2[1+8\sin^2\alpha+\frac{64}{27}\cos^4\alpha-12\epsilon^2(1-\frac{16}{27}\cos^2\alpha)]
\nonumber \\
\frac{\kappa_2}{C_0^2}\, &=&\, \frac{64\pi^2}{135}(\cos^4\alpha +3\,\epsilon^4
\frac{\cos^2{\alpha}}{\sin^4\alpha}).
\label{kasqt3}
\end{eqnarray}
The expression for $\kappa_2$ explains point (iv) regarding
$\kappa_\ell$. Interestingly, a residual $\kappa_2$ remains even for
very large space ($\epsilon\rightarrow 0$).

Preferred directions and statistically anisotropic CMB anisotropy have
been discussed in literature~\cite{fer_mag97bun_scot00}.  We compute
SI violation of the CMB anisotropy due to cosmic topology quantified
in terms of the recently proposed $\kappa_\ell$
spectrum~\cite{us_apj}. We interpret generic features of $\kappa_\ell$
spectrum arising from the shape of the Dirichlet domain as well as
through analytic results for leading order approximation to the
correlation.  These results help interpret observed non-zero
$\kappa_\ell$ and discriminate between different topologies of the
universe. When CMB anisotropy is multiply imaged, the $\kappa_\ell$
spectrum corresponds to a correlation pattern of matched pairs of
circles~\cite{circles}.  $\kappa_\ell$ have an advantage of being
insensitive to the overall orientation of the correlation features.
Also $\kappa_\ell$ are sensitive to SI violation even when CMB is not
multiply imaged.  The $A^{\ell M}_{ll^\prime}$ signature, which was
not discussed here, contains more details of the SI
violation~\cite{us_inprep}.

We study a new measure of SI violation.  A companion paper describes
the estimation of $\kappa_\ell$ from a CMB map~\cite{us_apj}.  Before
ascribing the detected breakdown of statistical anisotropy to
cosmological or astrophysical effects, one must carefully account for
and model out other mundane sources of statistical anisotropy in real
data, such as, incomplete and non-uniform sky coverage, beam
anisotropy, foreground residuals and statistically anisotropic noise.
These observational artifacts will be discussed in future
publications.

\acknowledgments TS acknowledges fruitful discussions with J.R.Bond,
D.  Pogosyan, G. Starkman and J. Weeks.


\end{document}